\documentstyle[12pt]{article}

\begin{document}
\title{Chiral field theory of $0^{-+}$ glueball}
\author{Bing An Li\\
Department of Physics and Astronomy, University of Kentucky\\
Lexington, KY 40506, USA}

\maketitle
\begin{abstract}
A chiral field theory of $0^{-+}$ glueball is presented. By adding a $0^{-+}$ glueball field to a 
successful Lagrangian of chiral field theory of pseudoscalar, vector, and axial-vector mesons,
the Lagrangian of this theory is constructed. The couplings between the pseodoscalar glueball field and 
the mesons are via U(1) anomaly revealed. 
Quantitative study of the physical processes of the $0^{-+}$ glueball of $m=1.405\textrm{GeV}$ is presented. 
The theoretical predictions can be used to identify the $0^{-+}$ glueball.

\end{abstract}
\newpage
\section{Introduction}
It is known for a very long time that glueball is the solution of nonperturbative QCD and there are extensive study 
on pseudoscalar glueballs[1]. 
On the other hand, many candidates of 
$0^{++},\;0^{-+}$, and $2^{++}$ glueballs have been discovered[2]. However, identification of a glueball 
is still in question. In order to identify a glueball 
quantitative study of the physical processes of a glueball is urgently needed.
It is the attempt of this paper to present a chiral field theory which can be used to do systematic and 
quantitative study of the 
properties of the $0^{-+}$ glueball.
   
Both current algebra and Lattice QCD successfully use quark operators to study nonpertuebative hadron physics. 
Based on current algebra and QCD
we have proposed a chiral field theory of pseudoscalar, vector, and axial-vector mesons[3], in which quark operators 
are used to study meson physics.
The Lagrangian of quarks and mesons is constructed as
\begin{eqnarray}
{\cal L}_{1}=\bar{\psi}(x)(i\gamma\cdot\partial
+\gamma\cdot v+\gamma\cdot a\gamma_{5}
-mu(x))\psi(x)-\bar{\psi}M\psi\nonumber \\
+{1\over 2}m^{2}_{0}(\rho^{\mu}_{i}\rho_{\mu i}+
\omega^{\mu}\omega_{\mu}+a^{\mu}_{i}a_{\mu i}+f^{\mu}f_{\mu}
+K^{*a}_{\mu}\bar{K}^
{*a\mu}+K_{1}^{\mu}K_{1\mu}
+\phi_{\mu}\phi^{\mu}+f_{s}^{\mu}f_{s\mu})
\end{eqnarray}
where \(a_{\mu}=\tau_{i}a^{i}_{\mu}+\lambda_{a}K^{a}_{1\mu}
+({2\over 3}+{1\over \sqrt{3}}\lambda_{8})
f_{\mu}+({1\over 3}-{1\over \sqrt{3}}\lambda_{8})
f_{s\mu}\)(\(i=1,2,3\) and \(a=4,5,6,7\)),
\(v_{\mu}=\tau_{i}
\rho^{i}_{\mu}+\lambda_{a}K^{a}_{\mu}+
({2\over 3}+{1\over \sqrt{3}}\lambda_{8})
\omega_{\mu}+({1\over 3}-{1\over \sqrt{3}}\lambda_{8})
\phi_{\mu}\), 
\(u=exp\{i\gamma_{5}
(\tau_{i}\pi_{i}+
\lambda_{a}K^{a}+\lambda_8\eta_8+
{1\over\sqrt{3}}\eta_0)\}\), 
m is the constituent quark mass which originates in the quark condensate,
M is the matrix of the current quark mass, $m_0$ is a parameter.
In the limit, $m_q\rightarrow 0$, the theory (1) has global $U(3)_L\times U(3)_R$ symmetry. 
In this theory the meson fields are related to corresponding quark operators. For example,
at the tree level the vector and the axial-vector mesons are expressed as the quark 
operators
\[\rho^i_\mu=-{1\over m^2_0}\bar{\psi}\gamma_\mu\tau^i\psi,\]
\[a^i_\mu=-{1\over m^2_0}\bar{\psi}\gamma_\mu\gamma_5\tau^i\psi.\]
The pseudoscalar mesons are via the mechanism of the nonlinear $\sigma$ model introduced. 
Under this mechanism the introduction of the constituent quark mass is natural and it 
plays an essential role in this theory. The mesons
are bound states of quarks and they are not independent degrees of freedoms and 
the kinetic terms of the meson fields are generated by the quark loop diagrams. Integrating out the quark fields,
the Lagrangian of the meson fields is derived.
$N_c$ expansion is naturally embedded. The tree diagrams are at the leading
order and the loop diagrams of the mesons are at the higher orders.
Besides the $N_C$ expansion there are current quark mass and momentum expansions in this theory.  
The major features of nonperturbative QCD: $N_C$ expansion, quark condensate, and chiral symmetry are all included in 
this meson theory.
The masses of the pseudoscalar, the vector, and the axial-vector mesons are determined. 
The form factors of the pion and the kaons are calculated in both space-like and time-like regions. 
The widths of
strong, electromagnetic, and weak decays of the mesons are computed. 
$\pi - \pi$ and 
$\pi - K$ scatterings are studied. The Wess-Zumino-Witten
anomaly is revealed. ChPT is the low energy approximation of this theory. All the 10 coefficients of the ChPT 
are determined.
Meson physics is systematically studied.  
The pion decay constant and a universal coupling constant are the two parameters in most cases. 
The third parameter, the quark condensate, 
only appears in the masses of the pseudoscalar mesons. Theory agrees 
with the data very well [3].
The meson physics are successfully studied by 
expressing the meson fields as the quark operators. The Lagrangian (1) is not complete. There are other degrees of freedoms,
for instance, glueballs. It is known that Lattice QCD has used  
gluon operator to calculate glueball mass [4]. Following the manner of Eq. (1),
using gluon operator to construct an effective Lagrangian to study the physics of the $0^{-+}$ glueball is the attempt of 
this paper. This paper is organized as: 1) Introduction; 2) Chiral Lagrangian of $0^{-+}$ glueball and mesons; 
3) Mass mixing of the $0^{-+}$ glueball $\eta(1405)$ and the $\eta,\; \eta'$; 4) $\eta(1405)\rightarrow\gamma\gamma$ decay;
5) $\eta(1405)\rightarrow\gamma\rho,\;\gamma\omega,\;\gamma\phi$ decays;
6) Kinetic mixing of $\chi$ and $\eta_0$ fields; 7) $J/\psi\rightarrow\gamma \eta(1405)$ decay;
8) $\eta(1405)\rightarrow\rho\pi\pi$ decay; 9) $\eta(1405)\rightarrow a_0(980)\pi$ decay;
10) $\eta(1405)\rightarrow K^* K$ decay; 
11) Summary.

\section{Chiral Lagrangian of $0^{-+}$ glueball and mesons} 

As mentioned above, Lattice Gauge Theory has used the gluon operator, $F\tilde{F}$(in the continuum limit),
to calculate the mass of the pseudoscalar glueball by the quench approximation [4]. 
The meson theory is phenomenologically successful,
in which the mesons are coupled to the quark operators. The same approach is used to construct 
an effective Lagrangian of $0^{-+}$ glueball in this paper. This theory should be 
chiral symmetric in the limit, $m_q\rightarrow 0$ and the field of the $0^{-+}$ glueball $\chi$ can 
be expressed as the gluon
operator $F\tilde{F}$. Under the least coupling principle the effective Lagrangian is constructed as
\begin{equation}  
{\cal L}=-{1\over4}F^{a\mu\nu}F^{a}_{\mu\nu}+F^{a}_{\mu\nu}\tilde{F}^{a\mu\nu}\chi+{1\over2}G^2_\chi \chi\chi,
\end{equation}
where
\(\tilde{F}^{\mu\nu}=\epsilon^{\mu\nu\alpha\beta}F_{\alpha\beta}\),
$G_\chi$ is a mass-related parameter. 
In QCD glueball is a bound state of gluons, not an independent degree of freedom, therefore, 
there is no kinetic term for the glueball field $\chi$.
Using Eq. (2), at the three level the 
glueball field is expressed as the gluon operator
\begin{equation}
\chi=-{1\over G^2_\chi}F_{\mu\nu}\tilde{F}^{\mu\nu}.
\end{equation}

The relationship between the gluon operator $F\tilde{F}$ and the quark operators is found from the U(1) anomaly
\begin{equation}
\partial_\mu(\bar{\psi}\gamma_\mu\gamma_5\psi)=2i\bar{\psi}M\gamma_5\psi+\frac{3g^2_s}{(4\pi)^2}F_{\mu\nu}\tilde{F}^{\mu\nu}.
\end{equation}
Using Eq. (4), Eq. (2) is rewritten as
\begin{equation}
{\cal L}=-{1\over4}F^{a\mu\nu}F^{a}_{\mu\nu}-(\frac{3g^2_s}{(4\pi)^2})^{-1}\{\bar{\psi}\gamma_\mu\gamma_5\psi\partial_\mu\chi
+2i\bar{\psi}M\gamma_5\psi\chi\}+{1\over2}G^2_\chi \chi\chi.
\end{equation}
The constant $(\frac{3g^2_s}{(4\pi)^2})^{-1}$ can be absorbed by the $\chi$ field. By redefining 
the $\chi$ field and the parameter $G_{\chi}$, 
Eq. (5) is rewritten as 
\begin{equation}
{\cal L}=-{1\over4}F^{a\mu\nu}F^{a}_{\mu\nu}-\{\bar{\psi}\gamma_\mu\gamma_5\psi\partial_\mu\chi
+2i\bar{\psi}M\gamma_5\psi\chi\}+{1\over2}G^2_\chi \chi\chi.
\end{equation}
The same symbols of $\chi$ and $G_\chi$ are used. 
Eq. (6) is chiral symmetric in the limit, $m_q\rightarrow 0$. It is known that $g^2_s N_c\sim 1$ in the $N_C$ expansion and
the loop diagrams with gluon internal lines are at the higher orders in the $N_C$ expansion. Therefore, at the leading order in $N_C$
expansion the kinetic terms of gluon fields are decoupled from this theory. 

Adding the two terms 
\begin{equation}
{\cal L}_{2}=-\{\bar{\psi}\gamma_\mu\gamma_5\psi\partial_\mu\chi
+2i\bar{\psi}M\gamma_5\psi\chi\}+{1\over2}G^2_\chi \chi\chi
\end{equation}
to the Lagrangian of mesons (1), 
the Lagrangian including the glueball field $\chi$ and the meson fields is found to be
\begin{equation}
{\cal L}={\cal L}_{1}+{\cal L}_{2}.
\end{equation}
As shown above, the ways introducing the $0^{-+}$ mesons and the $0^{-+}$ glueball field to the theory (8) are very different.
The couplings between the quark operators and the $\eta_8,\;\eta_0$ are different from the coupling of
the $0^{-+}$ glueball. 
For $\eta_8,\;\eta_0$ there are two couplings [3]:
\begin{equation}
-{c\over g}{2\over f_\pi}\bar{\psi}\gamma_\mu\gamma_5\lambda\psi\partial_\mu(\eta, \eta'),\;\;
-im{2\over f_\pi}\bar{\psi}\gamma_5\lambda\psi(\eta,\eta')
\end{equation}
where $\lambda=\lambda_8$ for $\eta_8$ and $\lambda={\sqrt{2}\over\sqrt{3}}I$ for $\eta_0$ respectively, 
\(c={f^2_\pi\over2g m^2_\rho}\), 
and g is a universal coupling constant and it is determined by the decay rate of
$\rho\rightarrow e^+ e$. In the chiral limit, the coupling for the $0^{-+}$ glueball is obtained from Eq. (6)
\begin{equation}
-\bar{\psi}\gamma_\mu\gamma_5\psi\partial_\mu\chi
\end{equation} 
The differences between Eqs.(9,10) lead to different physical results 
for the $0^{-+}$ mesons and the $0^{-+}$ glueball. The different physical results which are presented in this paper
should be able to distinguish a pseudoscalar glueball from the pseudoscalar mesons. 

Integrating out the quark fields, the kinetic term of the $\chi$ field and the vertices between the $\chi$ field and 
other mesons are obtained from the Lagrangian (8). This procedure is equivalent to do one quark loop calculation.
In the chiral limit, using the coupling (10) 
the quark loop diagram
\begin{equation}
\langle\chi(p')|S|\chi(p)\rangle=-{1\over2}\int d^4 x d^4 y\langle|T\{\{\bar{\psi}(x)\gamma_\mu\gamma_5\psi(x)
\bar{\psi}(y)\gamma_\nu\gamma_5\psi(y)\rangle p'_\mu p_\nu e^{i(p'x-py)}
\end{equation}
is calculated to $O(p^2)$ and the kinetic term of the $\chi$ field is found to be
\begin{equation}
{3\over2}F^2{1\over2}\partial_\mu\chi\partial_\mu\chi 
\end{equation}
where \(F^2(1-{2c\over g})=f^2_\pi\) [3]. 
The normalized $\chi$ field is determined as
\begin{equation}
\chi\rightarrow\sqrt{2\over3}{1\over F}\chi.
\end{equation}
It is the same as what has been done in Ref. [3] the couplings between the mesons and the $\chi$ field can via 
the vertex (10) be derived from the Lagrangian (8). As a matter of fact, all the meson vertices obtained from the coupling 
$-{c\over g}{2\sqrt{2}\over f_\pi}{1\over\sqrt{3}}\bar{\psi}\gamma_\mu\gamma_5\psi\partial_\mu\eta_0$ can be found in Ref. [3].
Replacing ${c\over g}{2\sqrt{2}\over f_\pi}{1\over\sqrt{3}}\eta_0$ in these meson vertices by $\sqrt{2\over3}{1\over F}\chi$
all the vertices involving the $\chi$ field are obtained.

\section{Mass mixing of the $0^{-+}$ glueball $\eta(1405)$ and the $\eta,\; \eta'$}
The matrix elements of Eq. (4)
have been used in the studies of the mixing between $\eta,\;\eta'$ and $0^{-+}$ glueball [5]. 
In Ref. [5b] a solution for the pseudoscalar glueball mass around $(1.4 \pm 0.1) \textrm{GeV}$ is presented. 
The mass of the $\chi$ field is taken as an input in this study.
Besides $\eta,\;\eta'$ there are other $I^G(J^{PC})=0^+(0^{-+})$ pseudoscalars listed in Ref. [2]: $\eta(1295),\;\eta(1405),\;
\eta(1475),\;\eta(1760)$. In Ref. [6] a systematic phenomenological analysis about these pseuscalars is presented. 
The analysis concludes that the $\eta(1405)$ is a possible candidate
of the $0^{-+}$ glueball. In this paper 
the theory (8) is applied to do systematic and quantitative study of the physical processes of the possible glueball 
state $\eta(1405)$. The theoretical predictions can 
be used to decide whether the $\eta(1405)$ is indeed a $0^{-+}$ glueball.
The same can be done to other possible candidate of the pseudoscalar glueball.
 
The chiral field theory (8) is
applied to study the mixing of $\eta,\;\eta'$ and $\eta(1405)$ 
in this section. In this chiral theory (1) the pion, kaon, and $\eta$ are Goldstone bosons. 
In the leading order in the chiral expansion their masses are found to be [3]
\begin{eqnarray}
m^2_{\pi^+}=-{4\over f^2_\pi}{1\over3}\langle0|\bar{\psi}\psi|0\rangle(m_u+m_d),\nonumber \\
m^2_{K^+}=-{4\over f^2_\pi}{1\over3}\langle0|\bar{\psi}\psi|0\rangle(m_u+m_s),\nonumber \\ 
m^2_{K^0}=-{4\over f^2_\pi}{1\over3}\langle0|\bar{\psi}\psi|0\rangle(m_d+m_s).
\end{eqnarray}
To the first order in current quark masses, the following elements of the mass matrices are 
derived from Eqs. (1,14) 
\begin{eqnarray}
m^2_{\eta_8}=-{4\over f^2_\pi}{1\over3}\langle0|\bar{\psi}\psi|0\rangle{1\over3}(m_u+m_d+4m_s)
={1\over3}\{2(m^2_{K^+}+m^2_{K^0})-m^2_\pi\},\nonumber\\
m^2_{\eta_0}=-{4\over f^2_\pi}{1\over3}\langle0|\bar{\psi}\psi|0\rangle{2\over3}(m_u+m_d+m_s)
={1\over3}(m^2_{K^+}+m^2_{K^0}+m^2_\pi),\nonumber \\
\Delta m^2_{\eta_8\eta_0}={4\sqrt{2}\over9}{1\over f^2_\pi}{1\over3}\langle0|\bar{\psi}\psi|0\rangle(m_u+m_d-2m_s)
={\sqrt{2}\over9}(m^2_{K^+}+m^2_{K^0}
-2m^2_\pi),
\end{eqnarray}
\(m^2_{\eta_8}=0.3211\; \textrm{GeV}^2\;and\;m^2_{\eta_0}=0.1703\; \textrm{GeV}^2\). If there is no $0^{-+}$ glueball
the mass of $\eta'$ is determined to be
\begin{equation}
m^2_{\eta'}=m^2_{\eta_8}+m^2_{\eta_0}-m^2_{\eta}=0.1911\;\textrm{GeV}^2
\end{equation}
which is much smaller than the physical value $0.9178\;\textrm{GeV}^2$. This problem is 
known as U(1) anomaly [7]. The diagram of two gluon exchange 
leads to additional mass term for $m^2_{\eta_0}$, which is proportional to 
$\frac{g^2_s}{(4\pi)^2}\langle 0|F\tilde{F}|\eta_0\rangle $ [7]. In this study $m^2_{\eta_0}$ is taken as a parameter.
It is necessary to point out that the current quark mass expansion and the $N_C$ expansion are two independent expansions 
in this theory.
Using Eqs. (7,9,14), the mixing between the $\eta_8$ 
and the $\chi$ is found to be
\begin{equation}
\Delta m^2_{\chi\eta_8}=-{4\sqrt{2}\over9}{1\over f_\pi F}{1\over3}\langle0|\bar{\psi}\psi|0\rangle (m_u+m_d-2m_s)
=-{\sqrt{2}\over9}{f_\pi\over F}(m^2_{K^+}+m^2_{K^0}-2m^2_\pi).
\end{equation}
Three of the elements of the mass matrix, $m^2_{\eta_8},\;m^2_{\eta_8\eta_0},\;m^2_{\chi\eta_8}$ are determined to the first order in
the current quark masses.
Both the current quark masses and two gluon exchange contribute to the matrix element 
$\Delta m^2_{\chi\eta_0}\equiv\Delta_3$, which is taken as a parameter. 
$m^2_{\eta_0}$, $\Delta_3$, and $m^2_{\chi}$ are the three parameters of the 
mass matrix of $\eta_8,\;\eta_0,\;\chi$. $m_{\eta(1405)}$, $m_\eta,\;and \;m_{\eta'}$ are taken as inputs.
The equation 
\begin{equation}
m^2_{\eta_8}+m^2_{\eta_0}+m^2_{\chi}=m^2_{\eta}+m^2_{\eta'}+m^2_{\eta(1405)}
\end{equation}
is one of the three eigen value equations of the mass matrix. The other two eigen value equations are derived as
\begin{eqnarray}
\Delta^2_3+m^4_{\eta_0}-2.87m^2_{\eta_0}+1.77=0,\\
\Delta^2_3+m^4_{\eta_0}-2.88m^2_{\eta_0}+1.74-0.02527\Delta_3=0.
\end{eqnarray}
The difference between these two equations is very small. It is very interesting to notice that in the
chiral limit, $m^2_{\eta_8},\;\Delta m^2_{\eta_8\eta_0} (15),and\;\Delta m^2_{\chi\eta_8} (17)\rightarrow 0$. Therefore, in the 
limit, $m_q\rightarrow 0$, the $\eta_8,\;\eta_0,\;\chi$ mixing is reduced to the $\eta_0-\chi$ mixing. 
 The two eigen value equations of the mass 
matrix of $\eta_0$ and $\chi$ are found to be 
\begin{eqnarray}
m^2_{\eta_0}+m^2_{\chi}=m^2_{\eta'}+m^2_{\eta(1405)},\nonumber \\
\Delta^2_3+m^4_{\eta_0}-2.89m^2_{\eta_0}+1.81=0.
\end{eqnarray} 
Because of $m^2_{\eta_8}\approx m^2_\eta$ the first equation of Eqs. (21) is very close to Eq. (18) and Eq. (19,20) 
are reduced to Eq. (21) in the chiral limit.
Therefore, the difference between Eqs.(19,20) is caused by 
the current quark masses. The masses of the current quarks listed in Ref. [2] spread in a wide range: 
\(m_u=1.5\;to\;3.3\;\textrm{MeV}\), 
\(m_d=3.5\;to\;6.0\;\textrm{MeV}\), \(m_s=104^{+26}_{-34}\;\textrm{MeV}\). The numerical values of $m^2_{\eta_8}$, 
$\Delta m^2_{\eta_8\eta_0}$ (15), and $\Delta m^2_{\chi\eta_8}$ (17) are computed by using the mass formulas (14, 15, 17) which are 
at the first order in the expansion of the current quark masses.  
The effect of the current quark masses at the second order can be seen from the pion masses.
At the first order in the current quark masses
$m^2_{\pi^+}=m^2_{\pi^0}$. The mass difference of $\pi^+$ and $\pi^0$ is about $3.5\%$ of the the average of the pion mass. 
Nonzero $m^2_{\pi^+}-m^2_{\pi^0}$ is resulted in the second order of the current quark masses [8] 
(of course, the electromagnetic interactions too). 
There should be errors caused by the current quark masses at higher orders in those values (15, 17). 
On the other hand, 
Eqs. (19,20,21) show that the effect of the current quark masses on the mixing of $\eta_0-\chi$ is small. 
In the study of the physics of $\eta(1405)$ Eq. (19)
is taken into account. Obviously, the values of the current quark masses affect the $\eta$ meson more. The physics of the $\eta$
meson won't be studied in this paper.

Solving the eigen equations of the mass matrix of $\eta_8,\;\eta_0,\;\chi$, 
\begin{eqnarray}
\eta &=& a_1\eta_8+b_1\eta_0+c_1\chi,\nonumber
\end{eqnarray}
\begin{eqnarray}
a_1=\frac{m_2-0.3+1.2453\Delta_3}{((m_2-0.3+1.2453\Delta_3)^2+0.0523\Delta_3+0.2422m_2-0.1944)^{{1\over2}}},\nonumber
\end{eqnarray}
\begin{eqnarray}
b_1=\frac{0.07108+0.3679\Delta_3}{m_2-0.3+1.2453\Delta_3}a_1,\;
c_1=\frac{-0.3679m_2+0.199}{m_2-0.3+1.2453\Delta_3}a_1,\nonumber
\end{eqnarray}
\begin{eqnarray}
\eta'&=&a_2\eta_8+b_2\eta_0+c_2\chi,\nonumber
\end{eqnarray}
\begin{eqnarray}
a_2=\frac{m_2+1.2453\Delta_3-0.9172}{0.07108-10.4432}b_2,\;\;
c_2=\frac{10.4432m_2-9.49}{0.07108-10.4432}b_2,\nonumber
\end{eqnarray}
\begin{eqnarray}
b_2=\frac{0.07108-10.4432}{((0.07108-10.4432)^2+(10.4432m_2-9.49)^2+(m_2+1.2453\Delta_3-0.9172)^2},\nonumber
\end{eqnarray}
\begin{eqnarray}
\eta(1405)&=&a_3\eta_8+b_3\eta_0+c_3\chi,\nonumber
\end{eqnarray}
\begin{eqnarray}
a_3=\frac{0.043}{1.9771-m_2}(-1.5768+\Delta_3+0.7988m_2)c_3,\;
b_3=\frac{0.002454+\Delta_3}{1.9771-m_2}c_3,\nonumber\\
c_3=\{1+\frac{1}{1.9771-m_2}[(0.002454+\Delta_3)^2+0.001849(-1.5768+\Delta_3+0.7988m_2)^2]\}^{-{1\over2}}
\end{eqnarray}
are obtained,
where \(m_2=m^2_{\eta_0}\), $a_1,\;b_2,\;c_3$ are taken to be positive.

\section{$\eta(1405)\rightarrow\gamma\gamma$ decay}
There is one independent parameter left in Eqs. (22), 
which can be determined by the decay rate of $\eta'\rightarrow\gamma\gamma$.
The Vector Meson Dominance(VMD) is a natural result of this chiral field theory [3]. The decay of pseudoscalar 
to two photons is an
anomalous process. The couplings between $\eta_8,\;\eta_0$ 
and $\rho\rho,\;\omega\omega,\;\phi\phi$ are presented in Ref. [3]
\begin{eqnarray}
{\cal L}_{\eta_8 vv}=\frac{N_C}{(4\pi)^2}\frac{8}{\sqrt{3}g^2 f_\pi}\eta_8\epsilon^{\mu\nu\alpha\beta}
\{\partial_\mu\rho^i_\nu\partial_\alpha\rho^i_\beta
+\partial_\mu\omega_\nu\partial_\alpha\omega_\beta-2\partial_\mu\phi_\nu\partial_\alpha\phi_\beta\},\nonumber\\
{\cal L}_{\eta_0 vv}=\frac{N_C}{(4\pi)^2}\frac{8\sqrt{2}}{\sqrt{3}g^2 f_\pi}\eta_0\epsilon^{\mu\nu\alpha\beta}
\{\partial_\mu\rho^i_\nu\partial_\alpha\rho^i_\beta
+\partial_\mu\omega_\nu\partial_\alpha\omega_\beta+\partial_\mu\phi_\nu\partial_\alpha\phi_\beta\}
\end{eqnarray}
The VMD leads to following relationships
\begin{eqnarray}
\rho^0_\mu\rightarrow{1\over2}egA_\mu,\;
\omega_\mu\rightarrow{1\over6}egA_\mu,\;
\phi_\mu\rightarrow{-1\over3\sqrt{2}}egA_\mu.
\end{eqnarray}
The couplings
\begin{eqnarray}
{\cal L}_{\eta_8\gamma\gamma}=\frac{\alpha N_C}{4\pi}\frac{8}{\sqrt{3} f_\pi}\eta_8{1\over6}
\epsilon^{\mu\nu\alpha\beta}\partial_\mu A_\nu\partial_\alpha A_\beta,\nonumber\\
{\cal L}_{\eta_0\gamma\gamma}=\frac{\alpha N_C}{4\pi}\frac{8\sqrt{2}}{\sqrt{3} f_\pi}\eta_0{1\over3}
\epsilon^{\mu\nu\alpha\beta}\partial_\mu A_\nu\partial_\alpha A_\beta
\end{eqnarray}
are found from Eqs. (23,24).
The vertex ${\cal L}_{\chi vv}$ is determined by the couplings (8)
\begin{eqnarray}
-\sqrt{2\over3}{1\over F}\bar{\psi}\gamma_\mu\gamma_5\psi\partial_\mu\chi,\;
{1\over g}\bar{\psi}\tau^i\gamma_\mu\psi\rho^i_\mu, ,\;
{1\over g}\bar{\psi}\gamma_\mu\psi\omega_\mu, \;
-{\sqrt{2}\over g}\bar{s}\gamma_\mu s\phi_\mu,
\end{eqnarray}
where ${1\over g}$ and ${\sqrt{2}\over g}$ are the normalization factor of the fields $\rho,\;\omega$ and $\phi$ respectively, 
s is the strange quark field. 
The calculation (to the fourth orders in the covariant derivatives) shows that 
two terms are found from the triangle quark loop diagrams of the $\chi vv$ vertex 
\[{3\over4}{N_C\over g^2}{\sqrt{2}\over\sqrt{3}}{1\over F}{1\over (4\pi)^2}
\epsilon^{\mu\nu\alpha\beta}p_\mu(q_{1\nu}-q_{2\nu})e^{\lambda_1}_{\alpha}e^{\lambda_2}_{\beta}\]
and
\[-{3\over4}{N_C\over g^2}{\sqrt{2}\over\sqrt{3}}{1\over F}{1\over (4\pi)^2}
\epsilon^{\mu\nu\alpha\beta}p_\mu(q_{1\nu}-q_{2\nu})e^{\lambda_1}_{\alpha}e^{\lambda_2}_{\beta},\]
where $p,\;q_{1,2}$ are the momenta of the $\chi$ and the two vectors respectively, 
$e^{\lambda_{1,2}}_{\alpha}$ are the polarization vectors of the two vector fields respectively.
These two terms are canceled each other. Therefore, in the chiral limit the glueball $\chi$ component is not 
coupled to the vector-vector meson pairs, the vertex $\chi vv$ vanishes. 
The pure glueball field $\chi$ doesn't decay to two photons.
Besides the axial-vector couplings (the first equation of Eqs. (9)) the peudoscalars $\eta_8,\;\eta_0$ 
have the pseudoscalar couplings (the second equations of Eq. (9)). The peudoscalars $\eta_8,\;\eta_0$
are coupled to the vector-vector meson pairs and decay to two photons (23). 
This is a very important difference between the pure $0^{-+}$ glueball and the $\eta_8,\;\eta_0$ mesons.

In the chiral limit, only the $\eta_{8,0}$ components of the $\eta'$ meson contribute to 
the $\eta'\rightarrow\gamma\gamma$ decay. Using Eqs. (23), the decay width is found to be
\begin{equation}
\Gamma(\eta'\rightarrow\gamma\gamma)=\frac{\alpha^2}{16\pi^3}\frac{m^3_{\eta'}}{f^2_\pi}(2\sqrt{2\over3}b_2+{1\over\sqrt{3}}a_2)^2.
\end{equation}
where $f_\pi=0.182\;\textrm{GeV}$ is taken. The experimental data of 
$\Gamma(\eta'\rightarrow\gamma\gamma)$ is $4.31(1\pm0.13)\;\textrm{keV}$. By inputting
$\Gamma(\eta'\rightarrow\gamma\gamma)$ and using Eqs. (18,19,22),
\begin{equation}
m^2_{\eta_0}=1.25\; \textrm{GeV}^2,\;\;\Delta_3=0.51\;\textrm {GeV}^2,\;\;
m_{\chi}=1.28\;\textrm{GeV}.
\end{equation}
are determined.
Substituting the values of $m^2_{\eta_0},\;\Delta_3,\;and\; m_\chi$ into Eqs. (22)
the expressions of $\eta,\;\eta',\;\eta(1405)$ are found to be
\begin{eqnarray}
\eta=0.9742\eta_8+0.1593\eta_0-0.16\chi,\nonumber \\
\eta'=-0.1513\eta_8+0.8208\eta_0-0.551\chi,\nonumber \\
\eta(1405)=-0.003522\eta_8+0.5724\eta_0+0.8199\chi.
\end{eqnarray}
Eq. (29) shows that the $\eta'$ meson contains large component of the glueball and 
the the $\eta_0$ meson component in the $\eta(1405)$ is large too and the $\eta_8$ component in the $\eta(1405)$ is
negligibly small. The mixing between
the $\eta_0$ meson and the glueball $\chi$ is very strong. These results confirm that in the chiral limit
the $\eta_8-\eta_0-\chi$ mixing is reduced to the $\eta_0-\chi$ mixing. The effects of the current quark masses are small
on the $\eta_0-\chi$ mixing. 

The orthogonality between the expressions (29) show that the accuracy of the expression of $\eta$ 
is about $93\%$ and for $\eta'$ and $\eta(1405)$ the accuracyies are about $98\%$.
As mentioned above that the physical masses of $\eta,\;\eta',\;\eta(1405)$ are imposed as the eigen values of the
mass matrix of $\eta_8,\;\eta_0,\;\chi$ to determine the parameters of the mass matrix. On the other hand,
three of the six elements (15, 17) of the mass matrix have been already determined. The combination of these two factors lead to the
errors. The three elements $m^2_{\eta_8},
\;\Delta m^2_{\eta_8\eta_0},\;\Delta m^2_{\chi\eta_8}$ affect the $\eta$ meson most. They can be calculated to higher orders 
in the
current quark masses and the errors, especially, the error on $\eta$ will be reduced. 
The small error for $\eta(1405)$ indicates that the physical mass of the $\eta(1405)$ as one of the eigen values of the
mass matrix is acceptable and the $\eta(1405)$ fits the room of the $0^{-+}$ glueball well. 
Eq. (29) is applied to study the physics of the $\eta(1405)$ 
in this paper. 

Only the $\eta_0$ component of $\eta(1405)$ contributes to the $\eta(1405)\rightarrow\gamma\gamma$ decay and
and the glueball component $\chi$ is suppressed. Using Eqs. (23,29), 
\begin{equation}
\Gamma(\eta(1405)\rightarrow\gamma\gamma)=\frac{\alpha^2}{16\pi^3}\frac{m^3_{\eta(1405)}}{f^2_\pi}(2\sqrt{2\over3}b_3)^2
=4.55\;\textrm{keV}
\end{equation}
is predicted. In Eq. (30) the mass of the $\eta(1405)$ contributes a factor of 3.5 in comparison with 
$\Gamma(\eta'\rightarrow\gamma\gamma)$. The total width of the $\eta(1405)$ is $51.5 \pm 3.4\; \textrm{MeV}$ [2] and
\[B(\eta(1405)\rightarrow2\gamma)=0.87 (1 \pm 0.07)\times10^{-4}\] 
is obtained. This small branching ratio is consistent
with that $\eta(1405)$ has not been discovered in two photon collisions. 

The physical processes of the $\eta$ meson won't studied in this paper. However,
the $\eta\rightarrow2\gamma$ decay is taken as an example of the effects of the
current quark masses. Using Eqs. (23,29), 
$\Gamma(\eta\rightarrow\gamma\gamma)=0.361 \textrm{keV}$ is obtained. 
The data is $0.511 (1 \pm 0.06) \textrm{keV}$. The theoretical prediction
is lower than the data by about $40\%$.  
The coefficients of the expression of the $\eta$ (23)
are sensitive to the values of the current quark masses. For instance, if the $m^2_{\eta_8}$ is changed by 
about $10\%$, which is allowed by the date of the current quark masses presented in Ref. [2]
the agreement between the prediction of the $\Gamma(\eta\rightarrow\gamma\gamma)$ and the 
experimental value can be achieved.
 
\section{$\eta(1405)\rightarrow\gamma\rho,\;\gamma\omega,\;\gamma\phi$ decays}
The $\eta_8$ component of $\eta(1405)$ is ignored and the $\chi$ component doesn't contribute to the coupling of the $\eta(1405)vv$. 
The vertex of $\eta(1405)vv$ is determined by the quark component $\eta_0$ only. Using the VMD and Eqs. (22,23), 
\begin{eqnarray}
{\cal L}_{\eta(1405)\rho\gamma}=\frac{eN_C}{(4\pi)^2}\frac{8\sqrt{2}}{\sqrt{3}f_\pi}b_3\eta_0
\epsilon^{\mu\nu\alpha\beta}\partial_\mu\rho_\nu\partial\alpha A_\beta,\nonumber \\
{\cal L}_{\eta(1405)\omega\gamma}=\frac{eN_C}{(4\pi)^2}\frac{8\sqrt{2}}{3\sqrt{3}f_\pi}b_3\eta_0
\epsilon^{\mu\nu\alpha\beta}\partial_\mu\omega_\nu\partial\alpha A_\beta,\nonumber \\  
{\cal L}_{\eta(1405)\phi\gamma}=\frac{eN_C}{(4\pi)^2}\frac{16}{\sqrt{3}f_\pi}b_3\eta_0
\epsilon^{\mu\nu\alpha\beta}\partial_\mu\phi_\nu\partial\alpha A_\beta
\end{eqnarray}
are obtained, where \(b_3=0.5724\) (29).
The universal coupling constant $g=0.395$ is determined by the decay rate of $\rho\rightarrow e^+ e$. 
The decay widths are found to be
\begin{eqnarray}
\Gamma(\eta(1405)\rightarrow\rho\gamma)=(0.5724)^2\frac{3\alpha}{2\pi^4 g^2}{1\over f^2_\pi}k^3_{\rho},\nonumber \\
k_\rho={m_{\eta(1405)}\over 2}(1-{m^2_\rho\over m^2_{\eta(1405)}}),\nonumber \\
\Gamma(\eta(1405)\rightarrow\omega\gamma)={1\over 9}(0.5724)^2\frac{3\alpha}{2\pi^4 g^2}{1\over f^2_\pi}k^3_{\omega},\nonumber \\
k_\omega={m_{\eta(1405)}\over 2}(1-{m^2_\omega\over m^2_{\eta(1405)}}),\nonumber \\
\Gamma(\eta(1405)\rightarrow\phi\gamma)=({2\over9}0.5724)^2\frac{3\alpha}{2\pi^4 g^2}{1\over f^2_\pi}k^3_{\phi},\nonumber \\
k_\phi={m_{\eta(1405)}\over 2}(1-{m^2_\phi\over m^2_{\eta(1405)}}).
\end{eqnarray}
The numerate results are 
\begin{equation}
\Gamma(\eta(1405)\rightarrow\rho\gamma)=0.84 \textrm{MeV},\;\Gamma(\eta(1405)\rightarrow\omega\gamma)=90.3\textrm{kev},
\;\Gamma(\eta(1405)\rightarrow\phi\gamma)=58.2\textrm{kev}.
\end{equation}
The branching ratios of these three decay modes are 
\[1.63\times10^{-2} (1 \pm 0.07),\;1.75\times10^{-3}(1 \pm 0.07),\;1.13\times10^{-3} (1 \pm 0.07)\]
respectively.

\section{Kinetic mixing of $\chi$ and $\eta_0$}
Besides mass mixing between the pseudoscalar mesons and the $0^{-+}$ glueball studied in section 3, 
there is kinetic mixing between the $\eta_0$ meson and the $\chi$ glueball.
While the mass matrix is diagonalized and the physical states are determined, however, the matrix of 
the kinetic terms might not be diagonalized by these new physical states. The $\rho-\omega$ system is a good example. 
Eq. (1) shows that the mass matrix
of the $\rho$ and the $\omega$ mesons is diagonalized. The kinetic terms of the $\rho$ and the $\omega$ fields 
are generated by the quark loop
diagrams [3]. The $\rho$-fields are nonabelian gauge fields. 
The mixing between the kinetic terms of the $\rho^0$ and the $\omega$ fields is dynamically generated by the quark loops, 
which is determined by the mass difference of the current quark masses, $m_d - m_u$, and the electromagnetic interactions [3,9]
\[{\cal L}_{\rho-\omega}=\{-\frac{1}{4\pi^2g^2}{1\over m}(m_d-m_u)
+{1\over24}e^2 g^2\}(\partial_\mu\rho_\nu-\partial_\nu\rho_\mu)
(\partial_\mu\omega_\nu-\partial_\nu\omega_\mu).\]

In this chiral field theory while the kinetic terms of the $\eta_0$ field [3] and the $\chi$ field are generated 
by the quark loop diagrams
(12) the kinetic mixing, $\partial_\mu\eta_0\partial_\mu\chi$, is dynamically generated by the quark loops too. 
The coefficient of this mixing is determined by three vertices 
\begin{equation}
-\sqrt{{2\over3}}{1\over F}\bar{\psi}\gamma_\mu\gamma_5\psi\partial_\mu\chi - {1\over\sqrt{3}}{c\over g}{2\sqrt{2}\over f_\pi}
\bar{\psi}\gamma_\mu\gamma_5\psi\partial_\mu\eta_0
-im{1\over\sqrt{3}}{2\sqrt{2}\over f_\pi}\bar{\psi}\gamma_5\psi\eta_0.
\end{equation}
By calculating the S-matrix element $\langle\eta_0|S|\chi\rangle $, in the chiral limit the kinetic mixing is found to be
\begin{equation}
-(1-{2c\over g})^{{1\over2}}\partial_\mu\eta_0\partial_\mu\chi.
\end{equation}
This kinetic mixing cannot be refereed to the mass mixing. 
The amplitudes of $\eta(1405)\rightarrow\gamma\gamma, \gamma v$ have been calculated to the fourth orders in the covariant derivatives. 
At this order there is no 
contribution from the kinetic mixing term (35) to these processes. 

\section{$J/\psi\rightarrow\gamma \eta(1405)$ decay}
In pQCD the $J/\psi$ radiative decay is described as $J/\psi\rightarrow\gamma gg,\;gg\rightarrow meson$. 
Therefore, if the meson is strongly coupled
to two gluons it should be produced in $J/\psi$ radiative decay copiously. 
Both the $\eta'$ and the $\eta(1405)$ contain large components of the pure glueball state $\chi$. Therefore, 
large branching ratios of $J/\psi\rightarrow\gamma\eta',\;
\gamma \eta(1405)$ should be expected.

In Ref.[10] the decay width of the $J/\psi\rightarrow\gamma\chi$ is derived as
\begin{equation}
\Gamma(J/\psi\rightarrow\gamma\chi)=\frac{2^{11}}{81}\alpha
\alpha^2_s(m_c)\psi^2_J(0)f^2_G{1\over m^8_c}\frac{(1-{m^2\over m^2_J})^3}{\{1-2\frac{m^2}{m^2_J}
+{4m^2_c\over m^2_J}\}^2} \nonumber \\
\{2m^2_J-3m^2(1+{2m_c\over m_J})-16{m^3_c\over m_J}\}^2,
\end{equation}
where $\psi_J(0)$ is the wave function of the $J/\psi$ at the origin, $f_G$ is a parameter 
related to the glueball state $\chi$, m is the mass of the physical state which contains the $\chi$ 
state and it will be specified.
After replacing corresponding quantities in Eq. (36), $m_c\rightarrow m_b$, $m_J\rightarrow m_\Upsilon$, 
$Q_c={2\over3}\rightarrow Q_b=-{1\over3}$, the equation (36) has been applied to study $B(\Upsilon(1S)\rightarrow\gamma 
\eta'(\eta))$ [11] and very strong suppression by the mass of the b quark has been found in these processes.
The suppression leads to very small $B(\Upsilon(1S)\rightarrow\gamma
\eta'(\eta))$, which are consistent with the experimental upper limits of $B(\Upsilon(1S)\rightarrow\gamma
\eta'(\eta))$ [12].

The $\chi$ state of Eq. (36) is via both the mass mixing (29) and the kinetic mixing (35) related to the $\eta'$ and 
the $\eta(1405)$ respectively
\begin{eqnarray}
\langle\eta'|\chi(0)|0 \rangle = -0.551+0.8208(1-{2c\over g})^{{1\over2}}\frac{m^2_{\eta'}}{m^2_\chi-m^2_{\eta'}}=0.3044
,\nonumber \\
\langle \eta(1405)|\chi(0)|0 \rangle = 0.8199+0.0.5724(1-{2c\over g})^{{1\over2}}\frac{m^2_G}{m^2_\chi-m^2_{G}}=-1.7788.
\end{eqnarray}
In Eqs. (37) the widths of $\eta'$ and $\eta(1405)$ are ignored. Eqs.(37) show that the kinetic mixing 
(the second terms of Eqs. (37)) (35) plays an
essential role in those two matrix elements.
Inputting $\Gamma(J/\psi\rightarrow\gamma\eta')$, the parameter $f_G$ and $\psi^2_J(0)$ are canceled and the ratio
\begin{equation}
R=\frac{\Gamma(J/\psi\rightarrow\gamma \eta(1405))}{\Gamma(J/\psi\rightarrow\gamma \eta')}
\end{equation}
is calculated. The ratio (38) is very sensitive to the value of the mass of the c quark and
the ratio R increases with $m_c$ dramatically. 
This sensitivity has already been found in Refs. [10,11]. In Ref.[2] $m_c=1.27^{+0.07}_{-0.11}\textrm{GeV}$ is listed. 
$m_c=1.3 \textrm{GeV}$ is taken in Ref.[13] to fit the data of $J/\psi\rightarrow\gamma f_2(1270)$ 
and this value is consistent with the one listed in Ref. [2].
Inputting $\Gamma_{\eta'}=0.205\pm0.015 \textrm{MeV}$, $\Gamma_{\eta(1405)}=51.1\pm3.4\textrm{GeV}$, 
$B(J/\psi\rightarrow\gamma\eta')=(4.71\pm0.27)\times10^{-3}$, and $m_c=1.3\textrm{GeV}$ and using Eqs. (36,37), it is predicted 
\begin{equation}
B(J/\psi\rightarrow\gamma \eta(1405))=0.73 (1 \pm 0.2)\times10^{-3}.
\end{equation}

\section{$\eta(1405)\rightarrow\rho\pi\pi$ decay}
The decay $\eta(1405)\rightarrow\rho\pi\pi$ is an anomalous decay mode.
There are two subprocesses: (1) $\eta(1405)\rightarrow\rho\rho,\;\rho\rightarrow\pi\pi$, (2) $\eta(1405)\rightarrow\rho\pi\pi$
directly. Because $\chi\rightarrow\rho\rho$ vanishes only the $\eta_0\rightarrow\rho\rho$ (23) contributes to (1). The vertex $\rho\pi\pi$ can be found from Ref. [3]
\begin{eqnarray}
{\cal L}_{\rho\pi\pi} = {2\over g}f_{\rho\pi\pi}\epsilon_{ijk}\rho^i_\mu\pi^j\partial_\mu\pi^k,\nonumber \\
f_{\rho\pi\pi}= 1+ \frac{q^2}{2\pi^2 f^2_\pi}\{(1-{2c\over g})^2-4\pi^2 c^2\},
\end{eqnarray}
where q is the momenta of the $\rho$ meson and the $f_{\rho\pi\pi}$ is the intrinsic form factor generated by the quark loop and 
it is a prediction of this chiral field theory of mesons [3]. This intrinsic form factor makes the theoretical results of 
the form factors of pion and kaons and the decay widths of $\rho,\;K^*,\;\phi$ mesons in excellent agreements with the data [3].
The amplitude of the subprocess (1) is derived as
\begin{equation}
T^{(1)} = -0.5724\frac{4\sqrt{6}}{\pi^2 g^3 f_\pi}
\frac{f_{\rho\pi\pi}}{q^2-m^2_\rho+i\sqrt{q^2}\Gamma(q^2)}\epsilon^{\mu\nu\alpha\beta}k_\mu e^\lambda_{\nu} k_{1\alpha}k_{2\beta},
\end{equation}
where 0.5724 is the coefficient of the $\eta_0$ component of $\eta(1405)$ (29), \(q = k_1 + k_2\), $\Gamma(q^2)$ is the 
decay width of the $\rho$ meson. When \(q^2\rangle 4 m^2_\pi\) [3]
\begin{equation}
\Gamma(q^2) = \frac{f_{\rho\pi\pi}^2(q^2)}{12\pi g^2}\sqrt{q^2}(1-{4m^2_\pi\over q^2})^{{3\over2}}.
\end{equation}
The subprocess (2) is the decay mode without intermediate resonance. The vertex of this process is similar to 
$f_1\rightarrow\rho\pi\pi$ presented in Ref. [3] (Eq. (111) of Ref. [3]) and it is found to be
\begin{equation}
{\cal L}_{\chi\rho\pi\pi} = \frac{2\sqrt{2}}{\sqrt{3}g\pi^2 f^2_\pi F}(1-{4c\over g})\epsilon_{ijk}
\epsilon^{\mu\nu\alpha\beta}\partial_\mu \chi
\partial_\nu\pi^i\partial_\alpha\pi^j\rho^k_\beta.
\end{equation}
The amplitude of the subprocess (2) is derived from Eq. (43) 
\begin{equation}
T^{(2)} = 0.8199\frac{4\sqrt{2}}{\sqrt{3}g\pi^2}{1\over f^2_\pi}{1\over F}\epsilon^{\mu\nu\alpha\beta}p_\mu e^\lambda_\nu 
k_{1\alpha}k_{2\beta},
\end{equation}
where 0.8199 is the coefficient of the $\chi$ component of $\eta(1405)$ (29). Only the glueball component $\chi$ contributes 
to $T^{(2)}$. Adding the two amplitudes (41,44) together the amplitude of the process $\eta(1405)\rightarrow\rho^0\pi^+\pi^-$ 
is found and the decay width is computed 
\begin{equation}
\Gamma(\eta(1405)\rightarrow\rho^0\pi^+\pi^-) = 0.92\; \textrm{MeV}.
\end{equation}
$T^{(1)}$ dominates the decay. The branching ratio of this channel is about $1.8\%$. The small branching ratio is 
resulted in two factors: The invariant mass of $\pi\pi$ is less than $m_\rho$ and the phase space of three body decay is much smaller 
than the one of two body decay. There are other two decay modes
\begin{equation}
\Gamma(\eta(1405)\rightarrow\rho^+\pi^0\pi^-) = \Gamma(\eta(1405)\rightarrow\rho^-\pi^+\pi^0) 
= \Gamma(\eta(1405)\rightarrow\rho^0\pi^+\pi^-).
\end{equation}
The total branching ratio of $\eta(1405)\rightarrow\rho\pi\pi$ is $5.4\%$. 
 
\section{$\eta(1405)\rightarrow a_0(980)\pi$ decay}
Two body decay,  $\eta(1405)\rightarrow a_0(980)\pi$, should be the major decay mode of $\eta(1405)$.
In the Lagrangian (1) the isovector scalar field $a_0(980)$ is not included and in order to study this 
decay mode the $a_0(980)$ field must be introduced to the Lagrangian (1).
As mentioned in the section of introduction that a meson field is expressed as a quark operator in this theory. It is natural that
\begin{equation}
a_0(980) \sim \bar{\psi}\tau^i\psi a^i_0.
\end{equation}
The quantum numbers of $a_0(980)$ are \(J^{PC}=0^{++}\) and in the Lagrangian (1) there is already a term
\(-m\bar{\psi}u\psi\). The parameter m is originated in the quark condensate whose \(J^{PC}=0^{++}\) too.
It is proposed that the $a_0(980)$ field can be added
to the Lagrangian by modifying $-m\bar{\psi}u\psi$ to 
\begin{equation}
-{1\over2}\bar{\psi}\{(m+\tau^i a_0^i)u+u(m+\tau^i a_0^i)\}\psi.
\end{equation}
Of course a mass term 
\begin{equation}
{1\over 2}m^2_{a_0}a^i_0 a^i_0
\end{equation}
has to be introduced. At the tree level the combination of Eqs. (48,49) leads to
\begin{equation}
a^i_0=-{1\over m^2_{a_0}}\bar{\psi}\tau^i\psi
\end{equation}
Therefore, Eq. (47) is revealed from this scheme. 

The couplings between the mesons and the $a_0(980)$ can be derived from Eq. (48). Using the vertex 
\begin{equation}
{\cal L}=-\bar{\psi}\tau^i\psi a^i_0
\end{equation}
obtained from Eq. (48), the quark loop diagram of the S-matrix element $\langle a_0 |S| a_0 \rangle $ is calculated and  
the kinetic term of the $a_0$ field is found. The $a_0$ field is normalized to be
\begin{equation}
a_0\rightarrow \sqrt{2\over 3}{1\over g}(1-{1\over 3\pi^2 g^2})^{-{1\over2}} a_0.
\end{equation}
A mass term is generated from
the quark loop diagram. The mass of the $a_0$ field has to be redefined, which is taken as a parameter.

In this paper we focus on 
the decay $\eta(1405)\rightarrow a_0(980)\pi$ and the decays of $a_0(980)$ will be studied in another paper.
Ignoring the $\eta_8$ component of $\eta(1405)$, there are $\eta_0\rightarrow a_0\pi$ and $\chi\rightarrow a_0\pi$ two
processes. In this study the decay width of $\eta(1405)\rightarrow a_0\pi$ is calculated to the leading order in the momentum 
expansion. Because of the derivative coupling $-\sqrt{2\over3}{1\over F}\bar{\psi}\gamma_\mu\gamma_5\psi\partial_\mu\chi$
the $\chi\rightarrow a_0\pi$ channel is at the next leading order in the momentum expansion. Therefore, only 
the $\eta_0\rightarrow a_0\pi$ channel is taken into account. 
The vertices related to this channel are found from the vertex (48)
\begin{eqnarray}
{\cal L}=-i\frac{2\sqrt{2}}{\sqrt{3}f_\pi}m\bar{\psi}\gamma_5\psi\eta_0
-i\frac{2m}{f_\pi}\bar{\psi}\tau^i\gamma_5\psi\pi^i\nonumber \\
-\sqrt{2\over3}{1\over g}(1-{1\over3\pi^2 g^2})^{-{1\over2}}\bar{\psi}\tau^i\psi a^i_0
-i{2m\over f_\pi}\sqrt{2\over3}{1\over g}(1-{1\over3\pi^2 g^2})^{-{1\over2}}\bar{\psi}I\gamma_5\psi a^i_0\pi^i,
\end{eqnarray}
where I is a $2\times 2$ unit matrix.
To the leading order in the momentum expansion the amplitude obtained from these vertices (53) is found to be
\begin{equation}
T=-0.5724\frac{8\sqrt{2}}{\sqrt{3}f^2_\pi}{1\over g}(1-{1\over3\pi^2 g^2})^{-{1\over2}} 
\{{1\over3}\langle0|\bar{\psi}\psi|0\rangle + 3m^3 g^2\},
\end{equation}
where the coefficient 0.5724 is the component of the $\eta_0$ of the $\eta(1405)$ (29).
In the amplitude (54) the quark condensate is obtained from the vertex $\bar{\psi}I\gamma_5\psi a^i_0\pi^i$ which is derived from 
\begin{equation}
-{1\over2}i\bar{\psi}\gamma_5\{a_0 \pi+\pi a_0\}\psi
\end{equation}
of Eq. (48). The vertices, $\bar{\psi}\tau^i\gamma_5\psi\pi^i$ and $\bar{\psi}\tau^i\psi a^i_0$, which are obtained
from Eqs.(1,48), contribute to the term, $3m^3 g^2$, of Eq. (54). It is known that the quark 
condensate is negative. Therefore, there is cancellation between the two terms of the amplitude (54). The cancellation
makes the decay width narrower. The mechanism (48) introducing the $a_0$ field to this chiral field theory leads to the cancellation.
The decay width of $\eta(1405)\rightarrow a_0\pi$ is sensitive to the value of the quark condensate. 
\begin{equation}
{1\over3}\langle0|\bar{\psi}\psi|0\rangle = -(0.24)^3\; \textrm{GeV}
\end{equation}
is taken and it is close to the value used in Ref. [14]. The constituent quark mass m is determined in Ref. [3]
\begin{equation}
m^2={f^2_\pi\over 6g^2(1-{2c\over g})^2}.
\end{equation}
The total decay width of the three modes, $a^+_0\pi^-,\;a^-_0\pi^+,\;a^0_0\pi^0$ of $\eta(1405)\rightarrow a_0\pi$ 
is calculated to be 
\begin{equation}
\Gamma(\eta(1405)\rightarrow a_0\pi)=44\; \textrm{MeV}.
\end{equation}
The branching ratio $B(\eta(1405)\rightarrow a_0\pi)=86 (1 \pm 0.07)\%$.
Therefore, $\eta(1405)\rightarrow a_0\pi$ is the major decay mode of $\eta(1405)$.

The decays of $\eta(1405)\rightarrow K\bar{K}\pi,\; \eta\pi\pi,\;\eta'\pi\pi$ are more complicated, in which 
the couplings between $a_0$ and $K\bar{K},\;\eta\pi,\;\eta'\pi$; $f_0(980)$ and $\pi\pi,\;K\bar{K}$ ... are involved.
There are direct couplings(without intermediate state) too. The chiral field theory (8) can be applied to study these processes.
The study will be presented in the near future.

\section{$\eta(1405)\rightarrow K^*(890) K$ decay}
The decay mode $\eta(1405)\rightarrow K\bar{K}\pi$ has been found [2]. 
$\eta(1405)\rightarrow K^*(890) K$ is a possible
decay channel. This channel has normal parity. In order to study it the real part (with normal parity)
of the Lagrangian (1) is quoted from Ref. [3]
\begin{eqnarray}
\lefteqn{{\cal L}_{RE}=\frac{N_{c}}{(4\pi)^{2}}m^{2}{D\over 4}\Gamma
(2-{D\over 2})TrD_{\mu}UD^{\mu}U^{\dag}}\nonumber \\
 & &-{1\over 3}\frac{N_{c}}{(4\pi
)^{2}}{D\over 4}\Gamma(2-{D\over 2})Tr\{
v_{\mu\nu}v^{\mu\nu}+a_{\mu\nu}a^{\mu\nu}\}\nonumber \\
 & &+{i\over 2}\frac{N_{c}}{(4\pi)^{2}}Tr\{D_{\mu}UD_{\nu}U^{\dag}+
D_{\mu}U^{\dag}D_{\nu}U\}v^{\nu\mu}\nonumber \\
 & &+{i\over 2}\frac{N_{c}}{(4\pi)^{2}}Tr\{D_{\mu}U^{\dag}D_{\nu}U
 -D_{\mu}
UD_{\nu}U^{\dag}\}a^{\nu\mu} \nonumber \\
 & &+\frac{N_{c}}{6(4\pi)^{2}}TrD_{\mu}D_{\nu}UD^{\mu}D^{\nu}U^{\dag}
\nonumber \\
 & &-\frac{N_{c}}{12(4\pi)^{2}}Tr\{
D_{\mu}UD^{\mu}U^{\dag}D_{\nu}UD^{\nu}U^{\dag}
+D_{\mu}U^{\dag}D^{\mu}UD_{\nu}U^{\dag}D^{\nu}U
-D_{\mu}UD_{\nu}U^{\dag}D^{\mu}UD^{\nu}U^{\dag}\} \nonumber \\
 & &+{1\over 2}m^{2}_{0}(\rho^{\mu}_{i}\rho_{\mu i}+
\omega^{\mu}\omega_{\mu}+a^{\mu}_{i}a_{\mu i}+f^{\mu}f_{\mu}
\nonumber \\
 & &+K^{*a}_{\mu}\bar{K}^
{*a\mu}+K_{1}^{\mu}K_{1\mu} + \phi_{\mu}\phi^{\mu}+f_{s}^{\mu}f_{s\mu}),
\end{eqnarray}
where
\begin{eqnarray*}
D_{\mu}U=\partial_{\mu}U-i[v_{\mu}, U]+i\{a_{\mu}, U\},\\
D_{\mu}U^{\dag}=\partial_{\mu}U^{\dag}-i[v_{\mu}, U^{\dag}]-
i\{a_{\mu}, U^{\dag}\},\\
v_{\mu\nu}=\partial_{\mu}v_{\nu}-\partial_{\nu}v_{\mu}
-i[v_{\mu}, v_{\nu}]-i[a_{\mu}, a_{\nu}],\\
a_{\mu\nu}=\partial_{\mu}a_{\nu}-\partial_{\nu}a_{\mu}
-i[a_{\mu}, v_{\nu}]-i[v_{\mu}, a_{\nu}],\\
D_{\nu}D_{\mu}U=\partial_{\nu}(D_{\mu}U)-i[v_{\nu}, D_{\mu}U]
+i\{a_{\nu}, D_{\mu}U\},\\
D_{\nu}D_{\mu}U^{\dag}=\partial_{\nu}(D_{\mu}U^{\dag})
-i[v_{\nu}, D_{\mu}U^{\dag}]
-i\{a_{\nu}, D_{\mu}U^{\dag}\}.
\end{eqnarray*}
The $K_\mu$ field ($K^*$)is included in the $v_\mu$ and appears in either the commutators of 
$D_{\mu}U,\;D_{\mu}U^{\dag}$,
$D_{\nu}D_{\mu}U,\;D_{\nu}D_{\mu}U^{\dag}$ or $v_{\mu\nu}$.
The components of $\eta_0$ and $\chi$ are flavor singlets, therefore, only the component of $\eta_8$ which
is associated with $\lambda_8$ appears in the
commutator, $[\lambda_8, K_\mu]$. The vertex obtained from these commutators is
\begin{equation}
{\cal L}_{\eta(1405)K^*K} = a_3 cf_{ab8}\partial_\mu\eta_8 K^a_\mu K^b,
\end{equation}
where c is a constant determined by Eq. (59) and \(a_3=0.00352\) is from the component of the $\eta_8$ component of the
$\eta(1405)$ (29).
Obviously, the contribution of the vertex (60) to the decay $\eta(1405)\rightarrow K^* K$ is very small.

The field $\partial_\mu\chi$ can be included in the $a_\mu$ field. The term in Eq. (60),
\[Tr\{D_{\mu}UD_{\nu}U^{\dag}+ D_{\mu}U^{\dag}D_{\nu}U\}v^{\nu\mu},\]
needs a special attention. To the fourth order in derivatives
\begin{equation}
Tr\{D_{\mu}UD_{\nu}U^{\dag}+ D_{\mu}U^{\dag}D_{\nu}U\}v^{\nu\mu}=-8(1-{2c\over g})Tr\{\partial_\mu\chi
\partial_\nu K+\partial_\mu K \partial_\nu\chi\}
(\partial_\nu K_\mu-\partial_\mu K_\nu) =0.
\end{equation}
This theory predicts that the decay width of $\eta(1405)\rightarrow K^* K$ is very small.

The decay rate of $\eta(1405)\rightarrow K^* K$ is determined by the $\eta_8$ component of the $\eta(1405)$ (29).
It is shown in the section 3 that in the chiral limit the $\eta_8-\eta_0-\chi$ mixing is reduced to the $\eta_0-\chi$ mixing
and the $\eta_8$ component of the physical state $\eta(1405)$ vanishes. Therefore, in the chiral limit 
the chiral field theory (8) predicts that $\Gamma(\eta(1405)\rightarrow K^* K)=0$. At the next leading order in the chiral
expansion the $\eta(1405)$ contains the $\eta_8$ component and a small decay rate of $\eta(1405)\rightarrow K^* K$ is
expected. As mentioned in the section 3 the component $\eta_8$ of $\eta(1405)$, $a_3$, determined in this paper (29) is not accurate.
The accurate determination of $\Gamma(\eta(1405)\rightarrow K^* K)$ is beyond the scope of this paper.

A $0^{-+}$ resonance $\eta(1416)$ has been discovered in $\pi^- p\rightarrow K^+ K^-\pi^0 n$ at 18 $\textrm{GeV}$ [15].
The parameters of this state are determined as [15]
\[M=1416 \pm 4 \pm 2\; \textrm{MeV},\;\;\Gamma=42 \pm 10 \pm 9 \;\textrm{MeV}.\]
These values are close to $\eta(1405)'s$ [2]. The ratio of the branching ratios 
\begin{equation}
R=\frac{B(\eta(1416)\rightarrow K^* \bar{K} + c.c)}{B(\eta(1416)\rightarrow\rightarrow a_0\pi^0)}=0.084 \pm 0.024
\end{equation}
have been reported in Ref. [15]. The final state $a_0\pi$ has three states, therefore, the ratio (62) should be divided by 
3 and 
\begin{equation}
R= 0.028 \pm 0.008.
\end{equation}
Assuming the $\eta(1416)$ is the $\eta(1405)$, Eq.(63) shows that $\Gamma(\eta(1405)\rightarrow K^* K)$ is narrower than
$\Gamma(\eta(1405)\rightarrow a_0\pi)$ by two orders of magnitude. This result supports the prediction made by this chiral 
field theory.

\section{Summary}   
Based on a phenomenologically successful chiral meson theory and the U(1) anomaly a chiral field theory of $0^{-+}$ 
glueball has been constructed. Systematic and quantitative study of the properties of the 
candidate of the $0^{-+}$ glueball $\eta(1405)$ have been done by this theory. 
The study of $\eta_8,\;\eta_0,\;\chi$ mixing shows that the 
mass of $\eta(1405)$ fits the room of the pseudoscalar glueball well. The prediction of the small branching ratio 
of $\eta(1405)\rightarrow2\gamma$ is consistent with the fact that $\eta(1405)$ has not been found in two photon collisions.
The theory predicts that $\eta(1405)\rightarrow a_0(980)\pi$ is the major decay mode of $\eta(1405)$. 
A very small branching ratio of $\eta(1405)\rightarrow K^* K$ is predicted and the theory is consistent with the data.
The glueball component $\chi$ of the $\eta(1405)$ is the dominant contributor of
the $J/\psi\rightarrow\gamma\eta(1405)$ decay.
$B(J/\psi\rightarrow\gamma\eta(1405))$ is via the kinetic mixing predicted. 
The quark component $\eta_0$ of the $\eta(1405)$ is the dominant contributor of 
the decay $\eta(1405)\rightarrow\gamma\gamma,\;\gamma V,\;\rho\pi\pi,\;a_0\pi$. 
The glueball component $\chi$ of the $\eta(1405)$ 
is suppressed in these processes. 
This chiral field theory can be
applied to study other possible candidates of the $0^{-+}$ glueball by input their masses into the theory to make
quantitative predictions.


\begin{thebibliography}{50}
\bibitem{} D. L. Scharre et al. (Mark II Collaboration), Phys. Lett.
           97B, 329 (1980); C. Edwards et al. (Crystal Ball Collaboration), Phys. Rev.
           Lett. {\bf 49}, 259 (1982); 50, 219(E) (1983);
           J. F. Donoghue, K. Johnson, and B. A. Li, Phys. Lett. {\bf B 99}, 416 (1981);
           M. S. Chanowitz, Phys. Rev. Lett. {\bf 46}, 981 (1981); K. Ishikawa, Phys. Rev. Lett. {\bf 46}, 978 (1981);
           R. Lacaze and H. Navelet, Nucl. Phys. {\bf B 186}, 247 (1981);
           C. E. Carlson, J. J. Coyne, P. M. Fishbane, F. Gross, and S. Meshkov, Phys. Lett. {\bf B 98}, 110 (1981);
           J. M. Cornwall and A. Soni, Phys. Rev. {\bf D 29}, 1424, (1984) and {\bf D 32}, 764 (1985).
           M. Acciarri et al. (L3 Collaboration), Phys. Lett. {\bf B 501}, 1 (2001);
           D. M. Li, H. Yu, and S. S. Fang, Eur. Phys. J. {\bf C 28}, 335 (2003);
           L. Faddeev, A. J. Niemi, and U. Wiedner, Phys. Rev. {\bf D 70}, 114033 (2004);
           S. B. Gerasimov, M. Majewski, and V. A. Meshcheryakov, arXiv:0708.3762;
           N. Kochelev and D. P. Min, Phys. Lett. {\bf B 633}, 283 (2006);
           B. A. Li, Phys. Rev. {\bf D 74}, 034019 (2006); X. G. He, X. Q. Li, X. Liu, and J. P. Ma, Eur. Phys. J. {\bf C 49}, 
           731 (2007).
\bibitem{} C. Amsler, et al., Particle Data Group, Phys. Lett. {\bf B 667}, 1(2008).
\bibitem{} Bing An Li, Phys. Rev.{\bf D52}, 5165-5183(1995); {\bf D52}, 5184-5193(1995); see the review article, Proc. of intern. 
           Conf. on Flavor Phys., p.146, 5/31-6/6,2001, Zhang-Jia-Jie , China, edited by Y.L.Wu.
\bibitem{} Y. Chen et al., Phys. Rev. D 73, 014516 (2006); G. S. Bali et al. (UKQCD Collaboration), Phys. Lett. B
           309, 378 (1993); C. J. Morningstar and M. Peardon, Phys. Rev. D 60, 034509 (1999).
\bibitem{} (a)T. Feldmann, P. Kroll, and B. Stech, Phys. Rev. {\bf D 58}, 114006 (1998); Phys. Lett. {\bf B 449}, 339(1999);
           (b)Hai-Yang Cheng, Hsiang-Nan Li, and Keh-Fei Liu, Phys. Rev. {\bf D 79} 014024, (2009);
           (c)T. Gutsche, V. E. Lyubovitskij and M. C. Tichy, Phys. Rev. {\bf D 80}, 014014 (2009);
           F. Buisseret, V. Mathiau, C. Semay, Phys. Rev. {\bf D 80}, 074021 (2009);
           V. Mathiau, V. Vento, arXiv: 0910.0212; S. He, M. Huang, Q. S. Yan, Phys. rev. {\bf D 81}, 014003 (2010).   
\bibitem{} A. Masoni, C. Cicalo, and G. L. Usai, J. Phys. G: Part. Phys. {\bf 32} R293, (2006).
\bibitem{} E. Witten, Nucl. Phys. {\bf B 156} 269, (1979); G. Veneziano, Nucl. Phys. {\bf B 159}, 213 (1979).
\bibitem{} B. A.Li, Eur. Phys. J. {\bf A 10}, 347 (2001).
\bibitem{} B. A. Li and J. X. Wang, Phys. Lett.{\bf B 543}, 48 (2002); D. N. Gao, M. L. Yan, Eur. Phys. 
           J. {\bf A3}, 293 (1998). 
\bibitem{} H. Yu, B. A. Li, Q. X. Sheng, and M. M. Zhang, Phys. Energ. Fort. Phys. Nucl. {\bf 8}, 285 (1984).
\bibitem{} B. A. Li, Phys. Rev. {\bf D 77},097502 (2008)
\bibitem{} S.B.Athar et. al., CLEO Collaboration, Phys. Rev. {\bf D76}, 072003 (2007).  
\bibitem{} B. A. Li and Q. X. Shaen, Phys. Lett. {\bf B} (1982).
\bibitem{} H. X. Chen, A. Hosaka, and S. L. Zhu, Phys. Rev. {\bf D 76}, 094025 (2007); 
           V. Gimenez, V. Porretti, and J. Reyes, Eur. Phys. J. {bf C 41}, 535 (2005);
           M. A. Shifman, A. I. Vainshtein and V. I. Zakharov, Nucl. Phys., {\bf B 147}, 448 (1979). 
\bibitem{} G. S. Adams et al., B852 Collaboration, Phys. Lett. {\bf B 516}, 264 (2001).
\end{thebibliography}
\end{document}